\documentclass[prl,twocolumn,tightenlines,superscriptaddress,a4paper,lengthcheck]{revtex4-1}
\usepackage{amsfonts}
\usepackage{amsmath}
\usepackage{amssymb}
\usepackage{graphicx}
\usepackage{graphicx,amssymb,amsbsy,color}
\usepackage{bm}
\usepackage{bbold}

\begin{document}

\title{Turbulence scaling laws across the superfluid to supersolid transition}
\author{C.-H. Hsueh}
\affiliation{Department of Physics, National Taiwan Normal University, Taipei 11677, Taiwan}
\author{Y.-C. Tsai}
\affiliation{Department of Physics, National Changhua University of Education, Changhua
50058, Taiwan}
\affiliation{Department of Physics, National Taiwan Normal University, Taipei 11677, Taiwan}
\author{T.-L. Horng}
\affiliation{Department of Applied Mathematics, Feng Chia University, Taichung 40724, Taiwan}
\author{M. Tsubota}
\affiliation{Department of Physics, Osaka City University, Sugimoto 3-3-138, Sumiyoshi-ku, Osaka 558-8585, Japan}
\affiliation{The OCU Advanced Research Institute for Natural Science and Technology (OCARINA), Osaka, Japan}
\author{W. C. Wu}
\affiliation{Department of Physics, National Taiwan Normal University, Taipei 11677, Taiwan}
\date{\today}

\begin{abstract}
We investigate quantum turbulence in a two-dimensional trapped supersolid and demonstrate that both the
wave and vortex turbulence involve triple rather than dual cascades, as in a superfluid.
Because of the presence of a second gapless mode associated with translation symmetry breaking,
a new $k^{-13/3}$ scaling law is predicted to occur in the wave turbulence.
Simultaneous fast vortex-antivortex creation and annihilation in the interior
of the oscillating supersolid results in a $k^{-1}$ scaling law in the vortex turbulence.
Numerical simulations based on the Gross-Pitaevskii equation confirmed the predictions.
\end{abstract}
\pacs{03.75.-b, 67.80.-s, 32.80.Ee, 34.20.Cf}
\maketitle

Turbulence in a superfluid (SF), named quantum turbulence (QT), has recently attracted considerable
interest in both liquid helium \cite{Halperin_Tsubota_book,Tsubota13,Barenghi14,Vinen02,Skrbek12}
and atomic Bose-Einstein condensates (BEC)
\cite{White12,PhysRevA.69.053601,PhysRevA.76.045603,PhysRevA.77.063625,PhysRevA.80.051603,PhysRevA.80.023618,
PhysRevLett.103.045301,PhysRevLett.104.075301,PhysRevA.85.053641,
PhysRevA.86.053621,PhysRevA.91.053620}. Turbulence mostly associated with
{\em vortices} can be characterized by an energy spectrum following the
Kolmogorov $k^{-5/3}$ power law (also known as the K41 law) \cite{frisch1995tlk}.
This law describes the process wherein energy is conservatively transported through scales,
and this process is named an energy-cascade. Richardson \cite{QJ:QJ49704820311} provided
a useful presentation of the K41 law that indicated that far away from forcing and sink,
small eddies gain energy from the break-up of large eddies,
after which they themselves are broken into even smaller eddies, and so on.
Energy is thus transported conservatively to smaller scales. In a two-dimensional (2D) fluid,
in addition to energy, there exists another positive conserved quantity: enstrophy.
Thus it typically involves two scaling laws (dual cascades),
including the inverse energy-cascade $k^{-5/3}$ spectrum and the forward enstrophy-cascade
$k^{-3}$ \cite{doi:10.1063/1.1762301} (or $k^{-4}$ \cite{SAPM:SAPM1971504377})
spectrum, for vortex turbulence (VT) in a 2D fluid.

In addition to the hydrodynamic turbulence associated with vortices,
turbulence consisting of low-energy {\em waves}, named wave turbulence (WT), plays a
central role in a system of interactions. WT can be regarded as the
out-of-equilibrium statistical mechanics of random nonlinear waves.
As a general physical phenomenon, WT is observed in a vast range of nonlinear systems,
on quantum to astrophysical scales \cite{wave_turbulence}.
Similar to the occurrence of dual cascades in a 2D VT, WT can also exhibit dual cascades in a
similar system \cite{Kolmakov25032014}.
Using the four-wave-interaction scheme,
it was discovered that WT in an atomic BEC has an inverse waveaction-cascade
$k^{-1/3}$ and forward energy-cascade $k^{-1}$ spectra \cite{Kolmakov25032014}.
The latter,  named Kolmogorov-Zakharov (KZ) spectrum, is the analog of the Kolmogorov spectrum
of hydrodynamic vortex turbulence.
The dual-cascade behavior of BEC WT has a precise interpretation: the forward energy and
the inverse waveaction cascades correspond to the strongly nonequilibrated process
of evaporative cooling and the condensation process, respectively.
Recently turbulent cascade has been observed in an oscillating {\em uniform} BEC wherein
the density momentum distribution $\widetilde{n}\left(k\right)$ displays a $k^{-7/2}$ scaling
behavior (equivalently kinetic energy displays a $k^{-3/2}$ scaling
behavior) in the infrared regime associated with the acoustic wave turbulence of a compressible
superfluid \cite{doi:10.1038/nature20114}.

Supersolid (SS) is a state of matter that simultaneously possesses superfluidity and solidity
and in which both gauge and translational symmetries are broken.
The SS state has been verified in Rydberg-dressed BECs \cite{PhysRevLett.104.195302,PhysRevLett.104.223002,PhysRevLett.105.135301,PhysRevA.86.013619,PhysRevA.88.043646}
and such systems are excellent for investigating the quantum hydrodynamics
coupled to a coherent structure. Quantized vortex lattices in an atomic SS have been studied in
a couple of previous investigations \cite{PhysRevLett.108.265301,Sci.Rep.6.31801}.
Exploring how QT, including both VT and WT, behaves in such a system
across the SF-to-SS transition is an interesting research topic.
For VT, this Letter will demonstrate that an oscillating SS constantly generates vortices and
antivortices in the interior and that such a fast vortex-antivortex creation and annihilation
cycle can lead to a $k^{-1}$ scaling law \cite{PhysRevLett.86.3080}.

\begin{figure}[tb]
\includegraphics[height=1.53in,width=3.4in]{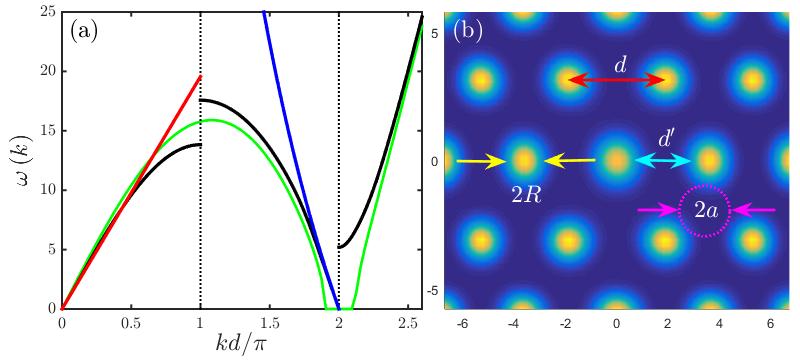}
\caption{(a) Excitation spectrum of an unstable uniform SF state (green curve) and true
periodic SS ground state [illustrated in (b), black curve] as a function of
$kd/\pi$ (where $d$ is the lattice constant). Red and blue curves display the asymptotic
behaviors of the black curve near $k=0$ and $k=(2\pi/d)^{-}$.
(b) 2D hexagonal SS lattice with lattice constant $d$, the gap between adjacent
SS droplets $d^\prime$, and droplet radius $R$;
the range $a$ in which the fluctuating vortices are mainly located is indicated.}
\label{fig1}
\end{figure}

For WT, the dispersion of elementary excitations should play a crucial role in determining
scaling laws. For a uniform SF condensate in the absence of an external potential,
the elementary excitation behaves according to
\begin{equation}
\hbar\omega\left(k\right)=\sqrt{\frac{\hbar^{2}k^{2}}{2m}\left[\frac{\hbar^{2}k^{2}}{2m}+
2n_{0}\widetilde{U}\left(k\right)\right]},
\label{dispersion}
\end{equation}
where $n_{0}$ is the uniform density and $\widetilde{U}\left(k\right)=\widetilde{U}\left(|\mathbf{k}|\right)$
is the Fourier transformation of the isotropic interaction function $U\left(\mathbf{r}\right)$.
If the interaction strength or mean density exceeds the critical value,
a modulation instability, named a roton instability, occurs \cite{PhysRevE.64.016612}.
The corresponding true ground state then undergoes a phase transition from a uniform SF
state to a periodic SS state. In Fig.~\ref{fig1}(a), the green curve plots the dispersion
(\ref{dispersion}) in an unstable SF state in which imaginary frequencies occur at a range
of wavevectors. By contrast, the black curve plots the excitation dispersion of the corresponding
true SS ground state in which a hexagonal lattice forms, as illustrated in Fig.~\ref{fig1}(b).
As displayed in Fig.~\ref{fig1}(a) in an extended zone scheme, there are {\em two} gapless modes
that appear near $k\rightarrow 0$ and $k\rightarrow (2\pi/d)^-$
with lattice constant $d$, respectively. (In an isotropic SF, $k=2\pi/d$ corresponds to the wavevector of the
roton before the instability occurred.) The asymptotic behaviors of the two gapless modes are
\begin{eqnarray}
\omega_1(k)&=& v_{1}k, ~~~~~~~~~~~~~~~~~~~~~~~~~~~~k\rightarrow 0 \nonumber\\
\omega_2(k)&=&\frac{\omega_{\rm{m}}^{2}}{v_{2}}k^{-1}-\omega_{\rm{m}},~~~~~~ k\rightarrow (2\pi/d)^-,
\label{omega12}
\end{eqnarray}
where $v_{1}=\omega_1^\prime(0)$, $v_{2}=\left|\omega_2^\prime(2\pi/d)\right|$, and the modulation
frequency $\omega_{\rm{m}}=2\pi v_{2}/d$. The two asymptotic dispersions
are indicated by the red and blue curves in Fig.~\ref{fig1}(a).

It is intuitively expected that the second gapless mode, which is the signature of a SS state,
may result in some new scaling law in WT.
Based on the theory of turbulence kinetics, the constant term
$\omega_{\rm{m}}$ in $\omega_2$ [see (\ref{omega12})] plays no role in determining the scaling law
because it will cancel out on both sides of the four-wave ($2\rightarrow2$) resonance condition.
Thus $\omega_2 \sim k^{-1}$ for $k\rightarrow (2\pi/d)^-$ so long as scaling law is concerned.
Consider wave dispersion that has an approximative form $\omega=\lambda k^{\alpha}$,
where $\lambda$ is a positive parameter and $\alpha$ is the power.
By dimension counting, the KZ spectra in an $N$-wave process will possess $k^{y}$ scaling laws for the
waveaction \cite{wave_turbulence}. Explicitly,
\begin{eqnarray}
  y &=& D-6+\frac{5-D+2\alpha\left(N-2\right)}{N-1},
\label{powerlaw}
\end{eqnarray}
where $D$ is the dimension \cite{wave_turbulence}.
With $D=2$, $N=4$, and $\alpha=-1$, we predict that a new scaling law with the power
\begin{eqnarray}
y=-{13\over 3}
\label{new_scaling}
\end{eqnarray}
will appear in the kinetic energy spectrum of a 2D SS that is associated with the breaking of translational symmetry
\cite{PhysRevLett.108.175301,PhysRevB.86.060510,PhysRevA.87.061602,PhysRevA.88.033618,Macrì2014}.

A numerical simulation was performed to investigate in detail the turbulence dynamics in a 2D SS. We employed
the Gross-Pitaevskii equation (GPE), and to study both the WT and VT, the kinetic energy was separated
into two mutually orthogonal parts: compressible and incompressible. Of most interest in this study
was whether a new $k^{-13/3}$ scaling law exists in the WT. The total energy of the current
interacting system can be written as the
summation of kinetic, potential, and interaction energies $E\left(t\right)=E_{\textrm{kin}}\left(t\right)+E_{\textrm{pot}}\left(t\right)+E_{\textrm{int}}\left(t\right)$, where
\begin{eqnarray}
E_{\textrm{kin}}\left(t\right)&=&\int\mathcal{E}_{\textrm{kin}}\left(\mathbf{r},t\right)d\mathbf{r}
=\int\frac{\hbar^{2}\left|\nabla\psi\left(\mathbf{r},t\right)\right|^{2}}{2m}d\mathbf{r},\nonumber\\
E_{\textrm{pot}}\left(t\right)&=&\int V_{\textrm{pot}}\left(r\right)\left|\psi\left(\mathbf{r},t\right)\right|^{2}d\mathbf{r},\nonumber\\
E_{\textrm{int}}\left(t\right)&=&\frac{1}{2}\int U\left(\bar{\mathbf{r}}\right)\left|\psi\left(\mathbf{r'},t\right)\right|^{2}
\left|\psi\left(\mathbf{r},t\right)\right|^{2}d\mathbf{r'}d\mathbf{r}.
\label{gpe}
\end{eqnarray}
Here, $V_{\textrm{pot}}\left(r\right)=m\omega_{\textrm{H}}^{2}r^{2}/2$ is the harmonic trapping
potential with a frequency $\omega_{\textrm{H}}$,
$U\left(\bar{\mathbf{r}}\right)=\mathcal{N}\widetilde{C}_{6}/\left(R_{\textrm{c}}^{6}+\bar{r}^{6}\right)$
is the soft-core interaction kernel with particle number $\mathcal{N}$,
$\widetilde{C}_{6}$ is the interaction strength, $R_{\textrm{c}}$ is the blockade radius, and $\bar{\mathbf{r}}\equiv\mathbf{r}-\mathbf{r'}$ is the relative position of two particles.
The order parameter $\psi$, satisfying the normalization
condition $\int\left|\psi\right|^{2}d\mathbf{r}=1$, is the condensate wave function.
Throughout this Letter, $R_{\textrm{c}}$ and $t_{0}\equiv mR_{\textrm{c}}^{2}/\hbar$ are used as
the units of length and time, respectively.  In our simulation, a SS condensate with Thomas-Fermi (TF)
radius $R_{\textrm{TF}}=6R_{\textrm{c}}$ was initially prepared in which circulation-3
vortex and antivortex were imprinted. Time evolution of the
wave function followed the GPE:  $i\hbar\partial\psi/\partial t=\delta E/\delta\psi$.

When applying the spectral scaling approach to study the turbulence in a trapped BEC,
it is useful to express the condensate wave function $\psi\left(\mathbf{r},t\right)$
in terms of Madelung transformation, $\psi\left(\mathbf{r},t\right)=\sqrt{n\left(\mathbf{r},t\right)}\exp\left[i\varphi\left(\mathbf{r},t\right)\right]$
with $n$ and $\varphi$ the density and phase, respectively.
The vector field $\sqrt{n}{\bf u}$ with the velocity $\mathbf{u}\equiv\left(\hbar/m\right)\nabla\varphi$
can then be decomposed into solenoidal and irrotational parts, or correspondingly,
incompressible and compressible parts
\cite{PhysRevLett.78.3896,doi:10.1063/1.869473,PhysRevLett.94.065302,doi:10.1143/JPSJ.74.3248}:
$\sqrt{n}\mathbf{u}=\left(\sqrt{n}\mathbf{u}\right)^{i}+\left(\sqrt{n}\mathbf{u}\right)^{c}$, where $\nabla\cdot\left(\sqrt{n}\mathbf{u}\right)^{i}=0$ and $\nabla\times\left(\sqrt{n}\mathbf{u}\right)^{c}=0$.
Because $\left(\sqrt{n}\mathbf{u}\right)^{i}$ and $\left(\sqrt{n}\mathbf{u}\right)^{c}$ are mutually orthogonal,
the kinetic energy density can also be decomposed accordingly: $\mathcal{E}_{\textrm{kin}}=\mathcal{E}_{\textrm{kin}}^{i}+\mathcal{E}_{\textrm{kin}}^{c}$, where $\mathcal{E}_{\textrm{kin}}^{i,c}=\left(m/2\right)|\left(\sqrt{n}\mathbf{u}\right)^{i,c}|^{2}$
correspond to the incompressible and compressible kinetic energy density, respectively.
Physically $\mathcal{E}_{\textrm{kin}}^{i}$ and $\mathcal{E}_{\textrm{kin}}^{c}$ correspond
to the kinetic energies of the swirls and the sound waves in a superflow, respectively.
To study the scaling laws, it is necessary to transform in $\mathbf{k}$ space: $E_{\textrm{kin}}(t)=E_{\textrm{kin}}^{i}(t)+E_{\textrm{kin}}^{c}(t)$ with $E_{\textrm{kin}}^{i,c}(t)=\int_{0}^{\infty}\widetilde{\mathcal{E}}_{1D}^{i,c}\left(k,t\right)dk$.
In a 2D space, $\widetilde{\mathcal{E}}_{1D}^{i,c}\left(k,t\right)$ is defined as
the angle-averaged kinetic-energy spectrum
\begin{equation}
\widetilde{\mathcal{E}}_{1D}^{i,c}\left(k,t\right)=\frac{k}{2}\int_{0}^{2\pi}d\phi_{k}
\widetilde{\mathcal{E}}_{\textrm{kin}}^{i,c}\left(\mathbf{k},t\right).
\label{kinetic_1D}
\end{equation}
The velocity field of the superflow may change constantly, but the energy spectrum
$\widetilde{\mathcal{E}}_{1D}^{i,c}$ takes on a stationary form as time increases.

\begin{figure}[tb]
\centering
\includegraphics[width=3.24in]{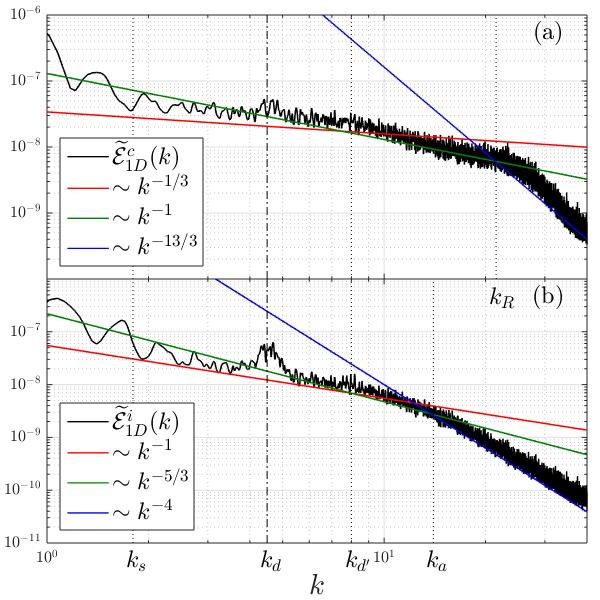}
\caption{(a) Compressible and (b) incompressible parts of the time-averaged kinetic energy spectrum
obtained near $t=150$.
Three scaling laws are identified in both cases, with the corresponding powers noted in the legend.
The critical wave vectors are identified as
$k_{s}=2\pi/s\approx 1.8$, $k_{d}=2\pi/d\approx 4.5$, $k_{d^\prime}=2\pi/d^\prime\approx 8$, $k_{a}=2\pi/a\approx 14$,
and $k_{R}=2\pi/R\approx 21.5$, corresponding to the radius of the eventual condensate $s$,
the 2D hexagonal SS lattice constant $d$, the gap between adjacent
SS droplets $d^\prime$, the range $a$ wherein fluctuating vortices mainly reside in a SS,
and the radius of the SS droplet€ $R$.}
\label{fig2}
\end{figure}

\begin{figure}[tb]
\begin{center}
\includegraphics[height=3.0in,width=3.5in]{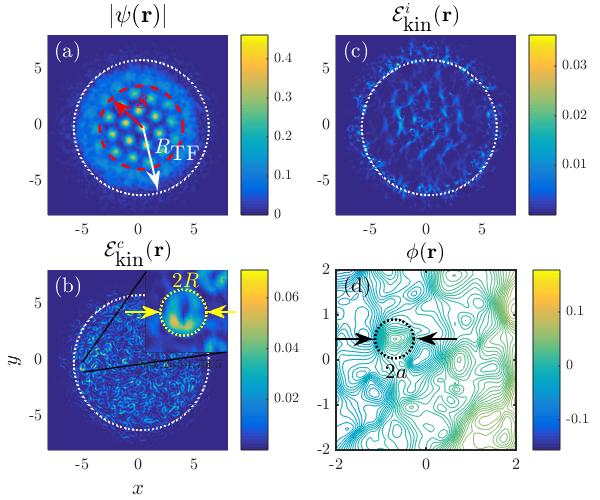}
\caption{(a) Normalized wave function $\psi$ obtained at $t=150$.
Dotted white and dashed red circles respectively indicate the initial and eventual sizes
of the condensate at $s\approx 3.5$ and $R_{\textrm{TF}}=6$.
(b) Compressible and (c) incompressible kinetic energy densities.
Inset in (b): size of SS droplet $R$ is
a crucial length scale for the new $k^{-13/3}$ scaling law in WT.
(d) Stream function $\phi\left(\mathbf{r}\right)$ of the corresponding vector field
with the length $a$ indicated (see text).}
\label{fig3}
\end{center}
\end{figure}

Figure~\ref{fig2} displays both the compressible (top) and incompressible (bottom) parts
of the time-averaged energy spectrum
$\widetilde{\mathcal{E}}_{1D}^{i,c}(k)$ in the vicinity of $t=150$,
which unambiguously exhibit scaling laws in both the
small-$k$ (large-$r$) and large-$k$ (small-$r$) regimes.
In a large-$k$ span such as is shown in Fig.~\ref{fig2},
both spectra indicate triple cascades rather than dual cascades, which normally occur in a uniform SF.
For the changes in the compressible energy, we identified [Fig.~\ref{fig2}(a)]
an inverse waveaction-cascade $k^{-1/3}$ spectrum
at $k_{s}<k<k_{d^\prime}$, which corresponds to the condensation process,
and a forward energy-cascade $k^{-1}$ spectrum at $k_{d^\prime}<k<k_{R}$, which
corresponds to energy transport away from the interior of the condensate
through splitting as small-scale sound waves.
At the connection between the two cascades, identified to be at the intersection
$k_{d^\prime}=2\pi/d^\prime\approx 8$ where
$d^\prime\approx0.8$ is the gap between adjacent SS droplets [see Fig.~\ref{fig1}(b)],
energy is input from the incompressible part. The lower limit of
the inverse waveaction-cascade spectrum $k^{-1/3}$ is consistent with the
eventual condensate size $s$ [see the dashed red circle in Fig.~\ref{fig3}(a)],
which corresponds to a wave number $k_{s}=2\pi/s\approx 1.8$ with $s\approx 3.5$.

Of most importance is the appearance of an additional inverse waveaction-cascade $k^{-13/3}$ spectrum
in the ultraviolet region that is the signature of the formation of a SS lattice. The lower limit of
this inverse waveaction-cascade spectrum at $k_{R}=2\pi/R\approx 21.5$
corresponds to the size of a SS droplet $R\approx0.3$.
At the connection of the forward energy-cascade spectrum $k^{-1}$ and the inverse
waveaction-cascade spectrum $k^{-13/3}$, energy is output to the incompressible part.
As mentioned earlier, typical low-energy wave dispersion is expressed as
$\omega=\lambda k^{\alpha}$ with a positive power
($\alpha>0$) and indicates a strictly increasing dispersion.
In the present SS system, however,
and due to the emergence of a roton, the wave dispersion can decrease due to a negative $\alpha$.
This is why a distinct $k^{-13/3}$ scaling law occurs in the ultraviolet region.
In the context of WT, SF-to-SS transitions can thus be regarded as the occurrence
of an inverse waveaction-cascade. This waveaction-cascade
is different from that associated with the formation of condensate.
Thus, there are two inverse waveaction-cascades in the present system;
the first is a result of the condensate size, as indicated by the dashed red circle in Fig.~\ref{fig3}(a),
and the second is due to the SS droplet size.

For the first gapless (phonon) mode $\omega_1=v_1 k$ at $k\rightarrow 0$
[see Fig.~\ref{fig1}(a)], the corresponding energy
spectrum is anticipated to have a forward energy-cascade spectrum $k^{-3/2}$ in the large-scale
(small-$k$) limit \cite{wave_turbulence}. This $k^{-3/2}$ spectrum was confirmed both by
numerically considering an infrared forcing \cite{PhysRevA.80.051603}
and by a recent experiment of a uniform system \cite{doi:10.1038/nature20114}.
When a SS forms, the $k$ range required to sustain the phonon mode dispersion
is relatively narrow \cite{PhysRevA.93.063605}, and consequently,
this $k^{-3/2}$ spectrum is not observed in the present system in the infrared regime
[see Fig.~\ref{fig2}(a)].

In the incompressible VT energy spectra displayed in Fig.~\ref{fig2}(b), both Kolmogorov $k^{-5/3}$
\cite{frisch1995tlk} and Saffman $k^{-4}$ \cite{SAPM:SAPM1971504377}
scaling laws can be identified. At the connection between the two
cascades at $k_{a}=2\pi/a\approx 14$ with $a\approx0.45$, energy is input from the compressible part.
When a SS lattice is present, length $a$ represents the range over which the fluctuating vortices
are mainly located [see Fig.~\ref{fig1}(b)].
The oscillation of a SS constantly creates vortices and antivortices in the interior,
and such a fast vortex-antivortex creation and annihilation cycle can result in a $k^{-1}$
scaling law in the infrared regime.
At the intersection of the spectra $k^{-1}$ and $k^{-5/3}$, energy is output to the compressible part.

Because the hexagonal lattice is one of the most critical characteristics of the present system,
the motion of the hexagonal structure produces not only vortices but also waves. Thus
lattice constant $d$ should be a crucial length scale in both the compressible and incompressible energies.
Taking into account the size of SS droplet, the gap between adjacent SS droplets $d^\prime=d-2R$ should be
another crucial length scale in both energies.
As was confirmed in Fig.~\ref{fig2}(a) and \ref{fig2}(b), $k_{d^\prime}$ is a critical wave vector in both the
compressible and incompressible spectra. More explicitly, $k_{d^\prime}$ represents a scale corresponding
to the energy source in the compressible part and the energy sink in the incompressible part.
Moreover, energy peak with wave vector $k_{d}$ occurs in both spectra.

Figure~\ref{fig3}(a) plots the normalized wave function at $t=150$ and displays both the initial and eventual
sizes ($R_{\rm TF}$ and $s$) of the condensate.
Fig.~\ref{fig3}(b) and \ref{fig3}(c) present the corresponding compressible and incompressible
kinetic energy densities. The magnification in the inset of Fig.~\ref{fig3}(b) emphasizes that
the size of a SS droplet ($R$) is an important (short) length scale for the additional
new inverse waveaction-cascade $k^{-13/3}$ spectrum in the ultraviolet region.
Plotted in Fig.~\ref{fig3}(d) is the stream function $\phi\left(\mathbf{r}\right)$ associated
with the velocity field $\sqrt{n}\mathbf{u}$ \cite{batchelor_2000}, which is the solution of Poisson's equation $\nabla^{2}\phi=\hat{\mathbf{z}}\cdot\mathbf{\Omega}$, where $\mathbf{\Omega}=\nabla\times\left(\sqrt{n}\mathbf{u}\right)$
is the 2D vorticity vector. The colors indicate the corresponding values of $\phi\left(\mathbf{r}\right)$, and the locations and structures of eddies can be easily recognized by families of closed streamlines.
From the definition of the vorticity vector $\mathbf{\Omega}$, we note that the singular and regular parts of $\mathbf{\Omega}$ are modulated by $\sqrt{n}$ and its gradient, respectively. As a result, the magnitude of the
vorticity $\mathbf{\Omega}$ of a vortex in high-density area is greater than that of a vortex in
low-density area and, with such reasoning, we state that an eddy is energetically ``larger"
if it contains more circumfluent particles.
Fig.~\ref{fig3}(d) clearly shows that the streams are mostly concentrated in a region of length $2a$,
which indicates that the fluctuating vortices mainly reside in this region.
The figure also reveals that the streamline bundles form a
(fluctuating) hexagonal lattice, which is consistent with the incompressible VT energy spectra displayed in
Fig.~\ref{fig3}(c).
By comparing the energy spectra of compressible and incompressible parts
[Fig.~\ref{fig3}(b) and \ref{fig3}(c)], the compressible and incompressible energies
are clearly observed to overlap not only in the border area but also in the interior.


In summary, QT including both VT and WT is investigated in a 2D trapped SS.
The SS state of an atomic BEC results from a roton instability, a type of modulation instability,
in a uniform SF. Because of the breaking of two symmetries, the SS possesses
two gapless modes.
The two gapless modes
behave differently near their zero-frequency wave vectors.
The second gapless mode associated with the translational symmetry breaking results in
a new $k^{-13/3}$ scaling law in the WT. Energy following the forward
cascade is eventually transferred into another form on small scales.
Because of the additional inverse $k^{-13/3}$ cascade in the present SF/SS system,
the kinetic energy is conservatively transported in circles between the compressible
and incompressible parts without any short-range dissipation.

CHH and YCT contributed equally to this work.
We thank K. Fujimoto for useful discussions.
Financial supports from MOST, Taiwan
(grant No. MOST 105-2112-M-003-005), JSPS KAKENHI (grant
numbers JP16H00807 and JP26400366), and NCTS of Taiwan are acknowledged.


\begin{thebibliography}{43}%
\makeatletter
\providecommand \@ifxundefined [1]{%
 \@ifx{#1\undefined}
}%
\providecommand \@ifnum [1]{%
 \ifnum #1\expandafter \@firstoftwo
 \else \expandafter \@secondoftwo
 \fi
}%
\providecommand \@ifx [1]{%
 \ifx #1\expandafter \@firstoftwo
 \else \expandafter \@secondoftwo
 \fi
}%
\providecommand \natexlab [1]{#1}%
\providecommand \enquote  [1]{``#1''}%
\providecommand \bibnamefont  [1]{#1}%
\providecommand \bibfnamefont [1]{#1}%
\providecommand \citenamefont [1]{#1}%
\providecommand \href@noop [0]{\@secondoftwo}%
\providecommand \href [0]{\begingroup \@sanitize@url \@href}%
\providecommand \@href[1]{\@@startlink{#1}\@@href}%
\providecommand \@@href[1]{\endgroup#1\@@endlink}%
\providecommand \@sanitize@url [0]{\catcode `\\12\catcode `\$12\catcode
  `\&12\catcode `\#12\catcode `\^12\catcode `\_12\catcode `\%12\relax}%
\providecommand \@@startlink[1]{}%
\providecommand \@@endlink[0]{}%
\providecommand \url  [0]{\begingroup\@sanitize@url \@url }%
\providecommand \@url [1]{\endgroup\@href {#1}{\urlprefix }}%
\providecommand \urlprefix  [0]{URL }%
\providecommand \Eprint [0]{\href }%
\providecommand \doibase [0]{http://dx.doi.org/}%
\providecommand \selectlanguage [0]{\@gobble}%
\providecommand \bibinfo  [0]{\@secondoftwo}%
\providecommand \bibfield  [0]{\@secondoftwo}%
\providecommand \translation [1]{[#1]}%
\providecommand \BibitemOpen [0]{}%
\providecommand \bibitemStop [0]{}%
\providecommand \bibitemNoStop [0]{.\EOS\space}%
\providecommand \EOS [0]{\spacefactor3000\relax}%
\providecommand \BibitemShut  [1]{\csname bibitem#1\endcsname}%
\let\auto@bib@innerbib\@empty
\bibitem [{\citenamefont {{W. P. Halperin and M.
  Tsubota}}(2009)}]{Halperin_Tsubota_book}%
  \BibitemOpen
  \bibfield  {author} {\bibinfo {author} {\bibnamefont {{W. P. Halperin and M.
  Tsubota}}},\ }\href@noop {} {\emph {\bibinfo {title} {{\rm Eds.} Prog. Low
  Temp. Phys. Vol. 16}}}\ (\bibinfo  {publisher} {Elsevier, Amsterdam},\
  \bibinfo {year} {2009})\BibitemShut {NoStop}%
\bibitem [{\citenamefont {{M. Tsubota, M. Kobayashi, and H.
  Takeuchi}}()}]{Tsubota13}%
  \BibitemOpen
  \bibfield  {author} {\bibinfo {author} {\bibnamefont {{M. Tsubota, M.
  Kobayashi, and H. Takeuchi}}},\ }\href@noop {} {}\Eprint
  {http://arxiv.org/abs/Phys. Rep. {\bf 522}, 191 (2013)} {Phys. Rep. {\bf
  522}, 191 (2013)} \BibitemShut {NoStop}%
\bibitem [{\citenamefont {{C. F. Barenghi, L. Skrbek, and K. P.
  Sreenivasan}}()}]{Barenghi14}%
  \BibitemOpen
  \bibfield  {author} {\bibinfo {author} {\bibnamefont {{C. F. Barenghi, L.
  Skrbek, and K. P. Sreenivasan}}},\ }\href@noop {} {}\Eprint
  {http://arxiv.org/abs/Proc. Natl. Acad. Sci. {\bf 111}, 4647 (2014)} {Proc.
  Natl. Acad. Sci. {\bf 111}, 4647 (2014)} \BibitemShut {NoStop}%
\bibitem [{\citenamefont {{W. F. Vinen and J. J. Niemela}}()}]{Vinen02}%
  \BibitemOpen
  \bibfield  {author} {\bibinfo {author} {\bibnamefont {{W. F. Vinen and J. J.
  Niemela}}},\ }\href@noop {} {}\Eprint {http://arxiv.org/abs/J. Low Temp.
  Phys. {\bf 128}, 167 (2002)} {J. Low Temp. Phys. {\bf 128}, 167 (2002)}
  \BibitemShut {NoStop}%
\bibitem [{\citenamefont {{L. Skrbek and K. P. Sreenivasan}}()}]{Skrbek12}%
  \BibitemOpen
  \bibfield  {author} {\bibinfo {author} {\bibnamefont {{L. Skrbek and K. P.
  Sreenivasan}}},\ }\href@noop {} {}\Eprint {http://arxiv.org/abs/Phys. Fluids
  {\bf 24}, 011301 (2012)} {Phys. Fluids {\bf 24}, 011301 (2012)} \BibitemShut
  {NoStop}%
\bibitem [{\citenamefont {{Angela C. White, Brian P. Anderson, and Vanderlei S.
  Bagnato}}()}]{White12}%
  \BibitemOpen
  \bibfield  {author} {\bibinfo {author} {\bibnamefont {{Angela C. White, Brian
  P. Anderson, and Vanderlei S. Bagnato}}},\ }\href@noop {} {}\Eprint
  {http://arxiv.org/abs/Proc. Natl. Acad. Sci. {\bf 111} 4719 (2014)} {Proc.
  Natl. Acad. Sci. {\bf 111} 4719 (2014)} \BibitemShut {NoStop}%
\bibitem [{\citenamefont {Berloff}(2004)}]{PhysRevA.69.053601}%
  \BibitemOpen
  \bibfield  {author} {\bibinfo {author} {\bibfnamefont {N.~G.}\ \bibnamefont
  {Berloff}},\ }\href {\doibase 10.1103/PhysRevA.69.053601} {\bibfield
  {journal} {\bibinfo  {journal} {Phys. Rev. A}\ }\textbf {\bibinfo {volume}
  {69}},\ \bibinfo {pages} {053601} (\bibinfo {year} {2004})}\BibitemShut
  {NoStop}%
\bibitem [{\citenamefont {Kobayashi}\ and\ \citenamefont
  {Tsubota}(2007)}]{PhysRevA.76.045603}%
  \BibitemOpen
  \bibfield  {author} {\bibinfo {author} {\bibfnamefont {M.}~\bibnamefont
  {Kobayashi}}\ and\ \bibinfo {author} {\bibfnamefont {M.}~\bibnamefont
  {Tsubota}},\ }\href {\doibase 10.1103/PhysRevA.76.045603} {\bibfield
  {journal} {\bibinfo  {journal} {Phys. Rev. A}\ }\textbf {\bibinfo {volume}
  {76}},\ \bibinfo {pages} {045603} (\bibinfo {year} {2007})}\BibitemShut
  {NoStop}%
\bibitem [{\citenamefont {Horng}\ \emph {et~al.}(2008)\citenamefont {Horng},
  \citenamefont {Hsueh},\ and\ \citenamefont {Gou}}]{PhysRevA.77.063625}%
  \BibitemOpen
  \bibfield  {author} {\bibinfo {author} {\bibfnamefont {T.-L.}\ \bibnamefont
  {Horng}}, \bibinfo {author} {\bibfnamefont {C.-H.}\ \bibnamefont {Hsueh}}, \
  and\ \bibinfo {author} {\bibfnamefont {S.-C.}\ \bibnamefont {Gou}},\ }\href
  {\doibase 10.1103/PhysRevA.77.063625} {\bibfield  {journal} {\bibinfo
  {journal} {Phys. Rev. A}\ }\textbf {\bibinfo {volume} {77}},\ \bibinfo
  {pages} {063625} (\bibinfo {year} {2008})}\BibitemShut {NoStop}%
\bibitem [{\citenamefont {Proment}\ \emph {et~al.}(2009)\citenamefont
  {Proment}, \citenamefont {Nazarenko},\ and\ \citenamefont
  {Onorato}}]{PhysRevA.80.051603}%
  \BibitemOpen
  \bibfield  {author} {\bibinfo {author} {\bibfnamefont {D.}~\bibnamefont
  {Proment}}, \bibinfo {author} {\bibfnamefont {S.}~\bibnamefont {Nazarenko}},
  \ and\ \bibinfo {author} {\bibfnamefont {M.}~\bibnamefont {Onorato}},\ }\href
  {\doibase 10.1103/PhysRevA.80.051603} {\bibfield  {journal} {\bibinfo
  {journal} {Phys. Rev. A}\ }\textbf {\bibinfo {volume} {80}},\ \bibinfo
  {pages} {051603} (\bibinfo {year} {2009})}\BibitemShut {NoStop}%
\bibitem [{\citenamefont {Horng}\ \emph {et~al.}(2009)\citenamefont {Horng},
  \citenamefont {Hsueh}, \citenamefont {Su}, \citenamefont {Kao},\ and\
  \citenamefont {Gou}}]{PhysRevA.80.023618}%
  \BibitemOpen
  \bibfield  {author} {\bibinfo {author} {\bibfnamefont {T.-L.}\ \bibnamefont
  {Horng}}, \bibinfo {author} {\bibfnamefont {C.-H.}\ \bibnamefont {Hsueh}},
  \bibinfo {author} {\bibfnamefont {S.-W.}\ \bibnamefont {Su}}, \bibinfo
  {author} {\bibfnamefont {Y.-M.}\ \bibnamefont {Kao}}, \ and\ \bibinfo
  {author} {\bibfnamefont {S.-C.}\ \bibnamefont {Gou}},\ }\href {\doibase
  10.1103/PhysRevA.80.023618} {\bibfield  {journal} {\bibinfo  {journal} {Phys.
  Rev. A}\ }\textbf {\bibinfo {volume} {80}},\ \bibinfo {pages} {023618}
  (\bibinfo {year} {2009})}\BibitemShut {NoStop}%
\bibitem [{\citenamefont {Henn}\ \emph {et~al.}(2009)\citenamefont {Henn},
  \citenamefont {Seman}, \citenamefont {Roati}, \citenamefont {Magalh\~aes},\
  and\ \citenamefont {Bagnato}}]{PhysRevLett.103.045301}%
  \BibitemOpen
  \bibfield  {author} {\bibinfo {author} {\bibfnamefont {E.~A.~L.}\
  \bibnamefont {Henn}}, \bibinfo {author} {\bibfnamefont {J.~A.}\ \bibnamefont
  {Seman}}, \bibinfo {author} {\bibfnamefont {G.}~\bibnamefont {Roati}},
  \bibinfo {author} {\bibfnamefont {K.~M.~F.}\ \bibnamefont {Magalh\~aes}}, \
  and\ \bibinfo {author} {\bibfnamefont {V.~S.}\ \bibnamefont {Bagnato}},\
  }\href {\doibase 10.1103/PhysRevLett.103.045301} {\bibfield  {journal}
  {\bibinfo  {journal} {Phys. Rev. Lett.}\ }\textbf {\bibinfo {volume} {103}},\
  \bibinfo {pages} {045301} (\bibinfo {year} {2009})}\BibitemShut {NoStop}%
\bibitem [{\citenamefont {White}\ \emph {et~al.}(2010)\citenamefont {White},
  \citenamefont {Barenghi}, \citenamefont {Proukakis}, \citenamefont {Youd},\
  and\ \citenamefont {Wacks}}]{PhysRevLett.104.075301}%
  \BibitemOpen
  \bibfield  {author} {\bibinfo {author} {\bibfnamefont {A.~C.}\ \bibnamefont
  {White}}, \bibinfo {author} {\bibfnamefont {C.~F.}\ \bibnamefont {Barenghi}},
  \bibinfo {author} {\bibfnamefont {N.~P.}\ \bibnamefont {Proukakis}}, \bibinfo
  {author} {\bibfnamefont {A.~J.}\ \bibnamefont {Youd}}, \ and\ \bibinfo
  {author} {\bibfnamefont {D.~H.}\ \bibnamefont {Wacks}},\ }\href {\doibase
  10.1103/PhysRevLett.104.075301} {\bibfield  {journal} {\bibinfo  {journal}
  {Phys. Rev. Lett.}\ }\textbf {\bibinfo {volume} {104}},\ \bibinfo {pages}
  {075301} (\bibinfo {year} {2010})}\BibitemShut {NoStop}%
\bibitem [{\citenamefont {Fujimoto}\ and\ \citenamefont
  {Tsubota}(2012)}]{PhysRevA.85.053641}%
  \BibitemOpen
  \bibfield  {author} {\bibinfo {author} {\bibfnamefont {K.}~\bibnamefont
  {Fujimoto}}\ and\ \bibinfo {author} {\bibfnamefont {M.}~\bibnamefont
  {Tsubota}},\ }\href {\doibase 10.1103/PhysRevA.85.053641} {\bibfield
  {journal} {\bibinfo  {journal} {Phys. Rev. A}\ }\textbf {\bibinfo {volume}
  {85}},\ \bibinfo {pages} {053641} (\bibinfo {year} {2012})}\BibitemShut
  {NoStop}%
\bibitem [{\citenamefont {Reeves}\ \emph {et~al.}(2012)\citenamefont {Reeves},
  \citenamefont {Anderson},\ and\ \citenamefont
  {Bradley}}]{PhysRevA.86.053621}%
  \BibitemOpen
  \bibfield  {author} {\bibinfo {author} {\bibfnamefont {M.~T.}\ \bibnamefont
  {Reeves}}, \bibinfo {author} {\bibfnamefont {B.~P.}\ \bibnamefont
  {Anderson}}, \ and\ \bibinfo {author} {\bibfnamefont {A.~S.}\ \bibnamefont
  {Bradley}},\ }\href {\doibase 10.1103/PhysRevA.86.053621} {\bibfield
  {journal} {\bibinfo  {journal} {Phys. Rev. A}\ }\textbf {\bibinfo {volume}
  {86}},\ \bibinfo {pages} {053621} (\bibinfo {year} {2012})}\BibitemShut
  {NoStop}%
\bibitem [{\citenamefont {Fujimoto}\ and\ \citenamefont
  {Tsubota}(2015)}]{PhysRevA.91.053620}%
  \BibitemOpen
  \bibfield  {author} {\bibinfo {author} {\bibfnamefont {K.}~\bibnamefont
  {Fujimoto}}\ and\ \bibinfo {author} {\bibfnamefont {M.}~\bibnamefont
  {Tsubota}},\ }\href {\doibase 10.1103/PhysRevA.91.053620} {\bibfield
  {journal} {\bibinfo  {journal} {Phys. Rev. A}\ }\textbf {\bibinfo {volume}
  {91}},\ \bibinfo {pages} {053620} (\bibinfo {year} {2015})}\BibitemShut
  {NoStop}%
\bibitem [{\citenamefont {Frisch}(1995)}]{frisch1995tlk}%
  \BibitemOpen
  \bibfield  {author} {\bibinfo {author} {\bibfnamefont {U.}~\bibnamefont
  {Frisch}},\ }\href@noop {} {\emph {\bibinfo {title} {Turbulence: The Legacy
  of AN Kolmogorov}}}\ (\bibinfo  {publisher} {Cambridge University Press},\
  \bibinfo {year} {1995})\BibitemShut {NoStop}%
\bibitem [{\citenamefont {Richardson}(1922)}]{QJ:QJ49704820311}%
  \BibitemOpen
  \bibfield  {author} {\bibinfo {author} {\bibfnamefont {L.~F.}\ \bibnamefont
  {Richardson}},\ }\href {\doibase 10.1002/qj.49704820311} {\bibfield
  {journal} {\bibinfo  {journal} {Quarterly Journal of the Royal Meteorological
  Society}\ }\textbf {\bibinfo {volume} {48}},\ \bibinfo {pages} {282}
  (\bibinfo {year} {1922})}\BibitemShut {NoStop}%
\bibitem [{\citenamefont {Kraichnan}(1967)}]{doi:10.1063/1.1762301}%
  \BibitemOpen
  \bibfield  {author} {\bibinfo {author} {\bibfnamefont {R.~H.}\ \bibnamefont
  {Kraichnan}},\ }\href {\doibase 10.1063/1.1762301} {\bibfield  {journal}
  {\bibinfo  {journal} {Phys. Fluids}\ }\textbf {\bibinfo {volume} {10}},\
  \bibinfo {pages} {1417} (\bibinfo {year} {1967})}\BibitemShut {NoStop}%
\bibitem [{\citenamefont {Saffman}(1971)}]{SAPM:SAPM1971504377}%
  \BibitemOpen
  \bibfield  {author} {\bibinfo {author} {\bibfnamefont {P.~G.}\ \bibnamefont
  {Saffman}},\ }\href {\doibase 10.1002/sapm1971504377} {\bibfield  {journal}
  {\bibinfo  {journal} {Studies in Applied Mathematics}\ }\textbf {\bibinfo
  {volume} {50}},\ \bibinfo {pages} {377} (\bibinfo {year} {1971})}\BibitemShut
  {NoStop}%
\bibitem [{\citenamefont {Nazarenko}(2011)}]{wave_turbulence}%
  \BibitemOpen
  \bibfield  {author} {\bibinfo {author} {\bibfnamefont {S.}~\bibnamefont
  {Nazarenko}},\ }\href@noop {} {\emph {\bibinfo {title} {Wave Turbulence}}}\
  (\bibinfo  {publisher} {Springer-Verlag Berlin Heidelberg},\ \bibinfo {year}
  {2011})\BibitemShut {NoStop}%
\bibitem [{\citenamefont {Kolmakov}\ \emph {et~al.}(2014)\citenamefont
  {Kolmakov}, \citenamefont {McClintock},\ and\ \citenamefont
  {Nazarenko}}]{Kolmakov25032014}%
  \BibitemOpen
  \bibfield  {author} {\bibinfo {author} {\bibfnamefont {G.~V.}\ \bibnamefont
  {Kolmakov}}, \bibinfo {author} {\bibfnamefont {P.~V.~E.}\ \bibnamefont
  {McClintock}}, \ and\ \bibinfo {author} {\bibfnamefont {S.~V.}\ \bibnamefont
  {Nazarenko}},\ }\href {\doibase 10.1073/pnas.1312575110} {\bibfield
  {journal} {\bibinfo  {journal} {Proc. Natl. Acad. Sci.}\ }\textbf {\bibinfo
  {volume} {111}},\ \bibinfo {pages} {4727} (\bibinfo {year}
  {2014})}\BibitemShut {NoStop}%
\bibitem [{\citenamefont {{N. Navon, A. L. Gaunt, R. P. Smith, and Z.
  Hadzibabic}}(2016)}]{doi:10.1038/nature20114}%
  \BibitemOpen
  \bibfield  {author} {\bibinfo {author} {\bibnamefont {{N. Navon, A. L. Gaunt,
  R. P. Smith, and Z. Hadzibabic}}},\ }\href {\doibase 10.1038/nature20114}
  {\bibfield  {journal} {\bibinfo  {journal} {Nature}\ }\textbf {\bibinfo
  {volume} {539}},\ \bibinfo {pages} {72} (\bibinfo {year} {2016})}\BibitemShut
  {NoStop}%
\bibitem [{\citenamefont {Henkel}\ \emph {et~al.}(2010)\citenamefont {Henkel},
  \citenamefont {Nath},\ and\ \citenamefont {Pohl}}]{PhysRevLett.104.195302}%
  \BibitemOpen
  \bibfield  {author} {\bibinfo {author} {\bibfnamefont {N.}~\bibnamefont
  {Henkel}}, \bibinfo {author} {\bibfnamefont {R.}~\bibnamefont {Nath}}, \ and\
  \bibinfo {author} {\bibfnamefont {T.}~\bibnamefont {Pohl}},\ }\href {\doibase
  10.1103/PhysRevLett.104.195302} {\bibfield  {journal} {\bibinfo  {journal}
  {Phys. Rev. Lett.}\ }\textbf {\bibinfo {volume} {104}},\ \bibinfo {pages}
  {195302} (\bibinfo {year} {2010})}\BibitemShut {NoStop}%
\bibitem [{\citenamefont {Pupillo}\ \emph {et~al.}(2010)\citenamefont
  {Pupillo}, \citenamefont {Micheli}, \citenamefont {Boninsegni}, \citenamefont
  {Lesanovsky},\ and\ \citenamefont {Zoller}}]{PhysRevLett.104.223002}%
  \BibitemOpen
  \bibfield  {author} {\bibinfo {author} {\bibfnamefont {G.}~\bibnamefont
  {Pupillo}}, \bibinfo {author} {\bibfnamefont {A.}~\bibnamefont {Micheli}},
  \bibinfo {author} {\bibfnamefont {M.}~\bibnamefont {Boninsegni}}, \bibinfo
  {author} {\bibfnamefont {I.}~\bibnamefont {Lesanovsky}}, \ and\ \bibinfo
  {author} {\bibfnamefont {P.}~\bibnamefont {Zoller}},\ }\href {\doibase
  10.1103/PhysRevLett.104.223002} {\bibfield  {journal} {\bibinfo  {journal}
  {Phys. Rev. Lett.}\ }\textbf {\bibinfo {volume} {104}},\ \bibinfo {pages}
  {223002} (\bibinfo {year} {2010})}\BibitemShut {NoStop}%
\bibitem [{\citenamefont {Cinti}\ \emph {et~al.}(2010)\citenamefont {Cinti},
  \citenamefont {Jain}, \citenamefont {Boninsegni}, \citenamefont {Micheli},
  \citenamefont {Zoller},\ and\ \citenamefont
  {Pupillo}}]{PhysRevLett.105.135301}%
  \BibitemOpen
  \bibfield  {author} {\bibinfo {author} {\bibfnamefont {F.}~\bibnamefont
  {Cinti}}, \bibinfo {author} {\bibfnamefont {P.}~\bibnamefont {Jain}},
  \bibinfo {author} {\bibfnamefont {M.}~\bibnamefont {Boninsegni}}, \bibinfo
  {author} {\bibfnamefont {A.}~\bibnamefont {Micheli}}, \bibinfo {author}
  {\bibfnamefont {P.}~\bibnamefont {Zoller}}, \ and\ \bibinfo {author}
  {\bibfnamefont {G.}~\bibnamefont {Pupillo}},\ }\href {\doibase
  10.1103/PhysRevLett.105.135301} {\bibfield  {journal} {\bibinfo  {journal}
  {Phys. Rev. Lett.}\ }\textbf {\bibinfo {volume} {105}},\ \bibinfo {pages}
  {135301} (\bibinfo {year} {2010})}\BibitemShut {NoStop}%
\bibitem [{\citenamefont {Hsueh}\ \emph {et~al.}(2012)\citenamefont {Hsueh},
  \citenamefont {Lin}, \citenamefont {Horng},\ and\ \citenamefont
  {Wu}}]{PhysRevA.86.013619}%
  \BibitemOpen
  \bibfield  {author} {\bibinfo {author} {\bibfnamefont {C.-H.}\ \bibnamefont
  {Hsueh}}, \bibinfo {author} {\bibfnamefont {T.-C.}\ \bibnamefont {Lin}},
  \bibinfo {author} {\bibfnamefont {T.-L.}\ \bibnamefont {Horng}}, \ and\
  \bibinfo {author} {\bibfnamefont {W.~C.}\ \bibnamefont {Wu}},\ }\href
  {\doibase 10.1103/PhysRevA.86.013619} {\bibfield  {journal} {\bibinfo
  {journal} {Phys. Rev. A}\ }\textbf {\bibinfo {volume} {86}},\ \bibinfo
  {pages} {013619} (\bibinfo {year} {2012})}\BibitemShut {NoStop}%
\bibitem [{\citenamefont {Hsueh}\ \emph {et~al.}(2013)\citenamefont {Hsueh},
  \citenamefont {Tsai}, \citenamefont {Wu}, \citenamefont {Chang},\ and\
  \citenamefont {Wu}}]{PhysRevA.88.043646}%
  \BibitemOpen
  \bibfield  {author} {\bibinfo {author} {\bibfnamefont {C.-H.}\ \bibnamefont
  {Hsueh}}, \bibinfo {author} {\bibfnamefont {Y.-C.}\ \bibnamefont {Tsai}},
  \bibinfo {author} {\bibfnamefont {K.-S.}\ \bibnamefont {Wu}}, \bibinfo
  {author} {\bibfnamefont {M.-S.}\ \bibnamefont {Chang}}, \ and\ \bibinfo
  {author} {\bibfnamefont {W.~C.}\ \bibnamefont {Wu}},\ }\href {\doibase
  10.1103/PhysRevA.88.043646} {\bibfield  {journal} {\bibinfo  {journal} {Phys.
  Rev. A}\ }\textbf {\bibinfo {volume} {88}},\ \bibinfo {pages} {043646}
  (\bibinfo {year} {2013})}\BibitemShut {NoStop}%
\bibitem [{\citenamefont {Henkel}\ \emph {et~al.}(2012)\citenamefont {Henkel},
  \citenamefont {Cinti}, \citenamefont {Jain}, \citenamefont {Pupillo},\ and\
  \citenamefont {Pohl}}]{PhysRevLett.108.265301}%
  \BibitemOpen
  \bibfield  {author} {\bibinfo {author} {\bibfnamefont {N.}~\bibnamefont
  {Henkel}}, \bibinfo {author} {\bibfnamefont {F.}~\bibnamefont {Cinti}},
  \bibinfo {author} {\bibfnamefont {P.}~\bibnamefont {Jain}}, \bibinfo {author}
  {\bibfnamefont {G.}~\bibnamefont {Pupillo}}, \ and\ \bibinfo {author}
  {\bibfnamefont {T.}~\bibnamefont {Pohl}},\ }\href {\doibase
  10.1103/PhysRevLett.108.265301} {\bibfield  {journal} {\bibinfo  {journal}
  {Phys. Rev. Lett.}\ }\textbf {\bibinfo {volume} {108}},\ \bibinfo {pages}
  {265301} (\bibinfo {year} {2012})}\BibitemShut {NoStop}%
\bibitem [{\citenamefont {Cheng}\ and\ \citenamefont
  {Jheng}(2016)}]{Sci.Rep.6.31801}%
  \BibitemOpen
  \bibfield  {author} {\bibinfo {author} {\bibfnamefont {S.-C.}\ \bibnamefont
  {Cheng}}\ and\ \bibinfo {author} {\bibfnamefont {S.-D.}\ \bibnamefont
  {Jheng}},\ }\href {\doibase 10.1038/srep31801} {\bibfield  {journal}
  {\bibinfo  {journal} {Scientific Reports}\ }\textbf {\bibinfo {volume} {6}},\
  \bibinfo {pages} {31801} (\bibinfo {year} {2016})}\BibitemShut {NoStop}%
\bibitem [{\citenamefont {Kivotides}\ \emph {et~al.}(2001)\citenamefont
  {Kivotides}, \citenamefont {Vassilicos}, \citenamefont {Samuels},\ and\
  \citenamefont {Barenghi}}]{PhysRevLett.86.3080}%
  \BibitemOpen
  \bibfield  {author} {\bibinfo {author} {\bibfnamefont {D.}~\bibnamefont
  {Kivotides}}, \bibinfo {author} {\bibfnamefont {J.~C.}\ \bibnamefont
  {Vassilicos}}, \bibinfo {author} {\bibfnamefont {D.~C.}\ \bibnamefont
  {Samuels}}, \ and\ \bibinfo {author} {\bibfnamefont {C.~F.}\ \bibnamefont
  {Barenghi}},\ }\href {\doibase 10.1103/PhysRevLett.86.3080} {\bibfield
  {journal} {\bibinfo  {journal} {Phys. Rev. Lett.}\ }\textbf {\bibinfo
  {volume} {86}},\ \bibinfo {pages} {3080} (\bibinfo {year}
  {2001})}\BibitemShut {NoStop}%
\bibitem [{\citenamefont {Krolikowski}\ \emph {et~al.}(2001)\citenamefont
  {Krolikowski}, \citenamefont {Bang}, \citenamefont {Rasmussen},\ and\
  \citenamefont {Wyller}}]{PhysRevE.64.016612}%
  \BibitemOpen
  \bibfield  {author} {\bibinfo {author} {\bibfnamefont {W.}~\bibnamefont
  {Krolikowski}}, \bibinfo {author} {\bibfnamefont {O.}~\bibnamefont {Bang}},
  \bibinfo {author} {\bibfnamefont {J.~J.}\ \bibnamefont {Rasmussen}}, \ and\
  \bibinfo {author} {\bibfnamefont {J.}~\bibnamefont {Wyller}},\ }\href
  {\doibase 10.1103/PhysRevE.64.016612} {\bibfield  {journal} {\bibinfo
  {journal} {Phys. Rev. E}\ }\textbf {\bibinfo {volume} {64}},\ \bibinfo
  {pages} {016612} (\bibinfo {year} {2001})}\BibitemShut {NoStop}%
\bibitem [{\citenamefont {Saccani}\ \emph {et~al.}(2012)\citenamefont
  {Saccani}, \citenamefont {Moroni},\ and\ \citenamefont
  {Boninsegni}}]{PhysRevLett.108.175301}%
  \BibitemOpen
  \bibfield  {author} {\bibinfo {author} {\bibfnamefont {S.}~\bibnamefont
  {Saccani}}, \bibinfo {author} {\bibfnamefont {S.}~\bibnamefont {Moroni}}, \
  and\ \bibinfo {author} {\bibfnamefont {M.}~\bibnamefont {Boninsegni}},\
  }\href {\doibase 10.1103/PhysRevLett.108.175301} {\bibfield  {journal}
  {\bibinfo  {journal} {Phys. Rev. Lett.}\ }\textbf {\bibinfo {volume} {108}},\
  \bibinfo {pages} {175301} (\bibinfo {year} {2012})}\BibitemShut {NoStop}%
\bibitem [{\citenamefont {Kunimi}\ and\ \citenamefont
  {Kato}(2012)}]{PhysRevB.86.060510}%
  \BibitemOpen
  \bibfield  {author} {\bibinfo {author} {\bibfnamefont {M.}~\bibnamefont
  {Kunimi}}\ and\ \bibinfo {author} {\bibfnamefont {Y.}~\bibnamefont {Kato}},\
  }\href {\doibase 10.1103/PhysRevB.86.060510} {\bibfield  {journal} {\bibinfo
  {journal} {Phys. Rev. B}\ }\textbf {\bibinfo {volume} {86}},\ \bibinfo
  {pages} {060510} (\bibinfo {year} {2012})}\BibitemShut {NoStop}%
\bibitem [{\citenamefont {Macr\`{\i}}\ \emph {et~al.}(2013)\citenamefont
  {Macr\`{\i}}, \citenamefont {Maucher}, \citenamefont {Cinti},\ and\
  \citenamefont {Pohl}}]{PhysRevA.87.061602}%
  \BibitemOpen
  \bibfield  {author} {\bibinfo {author} {\bibfnamefont {T.}~\bibnamefont
  {Macr\`{\i}}}, \bibinfo {author} {\bibfnamefont {F.}~\bibnamefont {Maucher}},
  \bibinfo {author} {\bibfnamefont {F.}~\bibnamefont {Cinti}}, \ and\ \bibinfo
  {author} {\bibfnamefont {T.}~\bibnamefont {Pohl}},\ }\href {\doibase
  10.1103/PhysRevA.87.061602} {\bibfield  {journal} {\bibinfo  {journal} {Phys.
  Rev. A}\ }\textbf {\bibinfo {volume} {87}},\ \bibinfo {pages} {061602}
  (\bibinfo {year} {2013})}\BibitemShut {NoStop}%
\bibitem [{\citenamefont {Ancilotto}\ \emph {et~al.}(2013)\citenamefont
  {Ancilotto}, \citenamefont {Rossi},\ and\ \citenamefont
  {Toigo}}]{PhysRevA.88.033618}%
  \BibitemOpen
  \bibfield  {author} {\bibinfo {author} {\bibfnamefont {F.}~\bibnamefont
  {Ancilotto}}, \bibinfo {author} {\bibfnamefont {M.}~\bibnamefont {Rossi}}, \
  and\ \bibinfo {author} {\bibfnamefont {F.}~\bibnamefont {Toigo}},\ }\href
  {\doibase 10.1103/PhysRevA.88.033618} {\bibfield  {journal} {\bibinfo
  {journal} {Phys. Rev. A}\ }\textbf {\bibinfo {volume} {88}},\ \bibinfo
  {pages} {033618} (\bibinfo {year} {2013})}\BibitemShut {NoStop}%
\bibitem [{\citenamefont {Macr{\`i}}\ \emph {et~al.}(2014)\citenamefont
  {Macr{\`i}}, \citenamefont {Saccani},\ and\ \citenamefont
  {Cinti}}]{Macrì2014}%
  \BibitemOpen
  \bibfield  {author} {\bibinfo {author} {\bibfnamefont {T.}~\bibnamefont
  {Macr{\`i}}}, \bibinfo {author} {\bibfnamefont {S.}~\bibnamefont {Saccani}},
  \ and\ \bibinfo {author} {\bibfnamefont {F.}~\bibnamefont {Cinti}},\ }\href
  {\doibase 10.1007/s10909-014-1192-7} {\bibfield  {journal} {\bibinfo
  {journal} {J. Low Temp. Phys.}\ }\textbf {\bibinfo {volume} {177}},\ \bibinfo
  {pages} {59} (\bibinfo {year} {2014})}\BibitemShut {NoStop}%
\bibitem [{\citenamefont {Nore}\ \emph
  {et~al.}(1997{\natexlab{a}})\citenamefont {Nore}, \citenamefont {Abid},\ and\
  \citenamefont {Brachet}}]{PhysRevLett.78.3896}%
  \BibitemOpen
  \bibfield  {author} {\bibinfo {author} {\bibfnamefont {C.}~\bibnamefont
  {Nore}}, \bibinfo {author} {\bibfnamefont {M.}~\bibnamefont {Abid}}, \ and\
  \bibinfo {author} {\bibfnamefont {M.~E.}\ \bibnamefont {Brachet}},\ }\href
  {\doibase 10.1103/PhysRevLett.78.3896} {\bibfield  {journal} {\bibinfo
  {journal} {Phys. Rev. Lett.}\ }\textbf {\bibinfo {volume} {78}},\ \bibinfo
  {pages} {3896} (\bibinfo {year} {1997}{\natexlab{a}})}\BibitemShut {NoStop}%
\bibitem [{\citenamefont {Nore}\ \emph
  {et~al.}(1997{\natexlab{b}})\citenamefont {Nore}, \citenamefont {Abid},\ and\
  \citenamefont {Brachet}}]{doi:10.1063/1.869473}%
  \BibitemOpen
  \bibfield  {author} {\bibinfo {author} {\bibfnamefont {C.}~\bibnamefont
  {Nore}}, \bibinfo {author} {\bibfnamefont {M.}~\bibnamefont {Abid}}, \ and\
  \bibinfo {author} {\bibfnamefont {M.~E.}\ \bibnamefont {Brachet}},\ }\href
  {\doibase 10.1063/1.869473} {\bibfield  {journal} {\bibinfo  {journal} {Phys.
  Fluids}\ }\textbf {\bibinfo {volume} {9}},\ \bibinfo {pages} {2644} (\bibinfo
  {year} {1997}{\natexlab{b}})}\BibitemShut {NoStop}%
\bibitem [{\citenamefont {Kobayashi}\ and\ \citenamefont
  {Tsubota}(2005{\natexlab{a}})}]{PhysRevLett.94.065302}%
  \BibitemOpen
  \bibfield  {author} {\bibinfo {author} {\bibfnamefont {M.}~\bibnamefont
  {Kobayashi}}\ and\ \bibinfo {author} {\bibfnamefont {M.}~\bibnamefont
  {Tsubota}},\ }\href {\doibase 10.1103/PhysRevLett.94.065302} {\bibfield
  {journal} {\bibinfo  {journal} {Phys. Rev. Lett.}\ }\textbf {\bibinfo
  {volume} {94}},\ \bibinfo {pages} {065302} (\bibinfo {year}
  {2005}{\natexlab{a}})}\BibitemShut {NoStop}%
\bibitem [{\citenamefont {Kobayashi}\ and\ \citenamefont
  {Tsubota}(2005{\natexlab{b}})}]{doi:10.1143/JPSJ.74.3248}%
  \BibitemOpen
  \bibfield  {author} {\bibinfo {author} {\bibfnamefont {M.}~\bibnamefont
  {Kobayashi}}\ and\ \bibinfo {author} {\bibfnamefont {M.}~\bibnamefont
  {Tsubota}},\ }\href {\doibase 10.1143/JPSJ.74.3248} {\bibfield  {journal}
  {\bibinfo  {journal} {J. Phys. Soc. Jpn.}\ }\textbf {\bibinfo {volume}
  {74}},\ \bibinfo {pages} {3248} (\bibinfo {year}
  {2005}{\natexlab{b}})}\BibitemShut {NoStop}%
\bibitem [{\citenamefont {Hsueh}\ \emph {et~al.}(2016)\citenamefont {Hsueh},
  \citenamefont {Tsai},\ and\ \citenamefont {Wu}}]{PhysRevA.93.063605}%
  \BibitemOpen
  \bibfield  {author} {\bibinfo {author} {\bibfnamefont {C.-H.}\ \bibnamefont
  {Hsueh}}, \bibinfo {author} {\bibfnamefont {Y.-C.}\ \bibnamefont {Tsai}}, \
  and\ \bibinfo {author} {\bibfnamefont {W.~C.}\ \bibnamefont {Wu}},\ }\href
  {\doibase 10.1103/PhysRevA.93.063605} {\bibfield  {journal} {\bibinfo
  {journal} {Phys. Rev. A}\ }\textbf {\bibinfo {volume} {93}},\ \bibinfo
  {pages} {063605} (\bibinfo {year} {2016})}\BibitemShut {NoStop}%
\bibitem [{\citenamefont {Batchelor}(2000)}]{batchelor_2000}%
  \BibitemOpen
  \bibfield  {author} {\bibinfo {author} {\bibfnamefont {G.~K.}\ \bibnamefont
  {Batchelor}},\ }\href {\doibase 10.1017/CBO9780511800955} {\emph {\bibinfo
  {title} {An Introduction to Fluid Dynamics}}},\ Cambridge Mathematical
  Library\ (\bibinfo  {publisher} {Cambridge University Press},\ \bibinfo
  {year} {2000})\BibitemShut {NoStop}%
\end{thebibliography}
%

\end{document}